# Charge Neutral MoS$_2$ Field Effect Transistors Through Oxygen Plasma Treatment

Rohan Dhall[1], Zhen Li[1], Ewa Kosmowska[2], Stephen B. Cronin[1]

[1]Ming Hsieh Department of Electrical Engineering, University of Southern California,

Los Angeles, 90089

[2]XEI Scientific, Redwood City, CA 94063

Lithographically fabricated MoS$_2$ field effect transistors suffer from several critical imperfections, including low sub-threshold swings, large turn-on gate voltages ($V_T$), and wide device-to-device variability. The large magnitude and variability of $V_T$ stems from unclean interfaces, trapped charges in the underlying substrate, and sulfur vacancies created during the mechanical exfoliation process. In this study, we demonstrate a simple and reliable oxygen plasma treatment, which mitigates the effects of unintentional doping created by surface defect cites, such as S vacancies, and surface contamination. This plasma treatment restores charge neutrality to the MoS$_2$ and shifts the threshold turn-on voltage towards 0V. Out of the 8 devices measured, all exhibit a shift of the FET turn-on voltage from an average of -18.7V to -0.9V. The oxygen plasma treatment passivates these defects, which reduces surface scattering, causing increased mobility and improved subthreshold swing. For as-prepared devices with low mobilities (~0.01cm$^2$/V·s), we observe up to a 190-fold increase in mobility after exposure to the oxygen plasma. Perhaps the most important aspect of this oxygen plasma treatment is that it reduces the device-to-device variability, which is a crucial factor in realizing any practical application of these devices.

**TOC FIGURE**

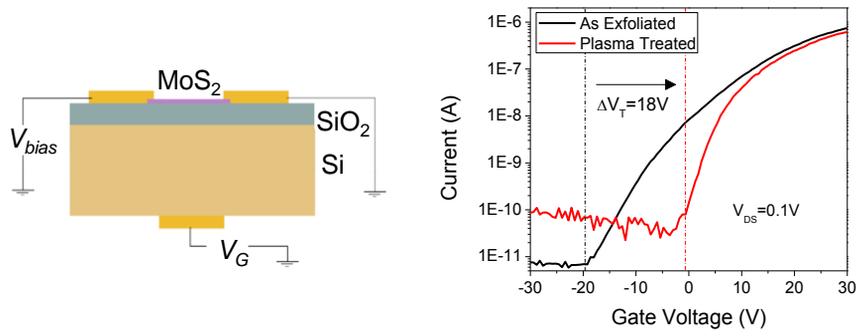

The unique properties of two dimensional materials have attracted a lot of interest for their potential applications in electronic and optoelectronic devices[1-5]. While the high carrier mobility of graphene[6] makes it an attractive material as a possible transparent electrode, it is of limited value for optoelectronics and digital logic due to its gapless electronic band structure. Transition metal dichalcogenides (TMDCs), such as $MoS_2$, $WSe_2$, and $WS_2$, are now being explored as possible alternatives for optoelectronic materials due to their finite band gaps, which lie in the visible range of the electromagnetic spectrum[7-10]. Monolayer $MoS_2$ is a direct band gap semiconductor[1, 5, 10], and hence, desirable for applications in optoelectronic devices, such as LEDs. Atomically thin $MoS_2$ field effect transistors (FETs) have also been demonstrated by several research groups. However, the utility of $MoS_2$ in high performance electronics is fundamentally limited. As pointed out by Yoon et al.[11], the carrier mobility in $MoS_2$ is limited by its relatively heavy electron mass ($0.45m_o$), and it is unlikely to be able to compete with state of the art III-V transistors. To further complicate matters, surface contamination and substrate interactions further degrade carrier mobilities in these two dimensional material systems. Nevertheless, considerable effort has gone into schemes to produce higher mobility $MoS_2$ FETs. For instance, Radisavljevic et al. showed improved field effect transistor (FET) mobilities in single layer $MoS_2$ transistors achieved by high temperature annealing and by using a dielectric coating of $HfO_2$ to passivate the $MoS_2$ surface[4]. Other studies have utilized hexagonal boron nitride encapsulation for enhanced mobility $MoS_2$ FET devices[12-16]. In our previous work, we reported an oxygen plasma treatment that improved the luminescence efficiency of few-layer $MoS_2$ by up to 20-fold due to an indirect-to-direct bandgap transition.[17] More recently, Ali Javey's group reported an air-stable, solution-based chemical treatment using an organic non-oxidizing superacid (bis(trifluoromethane) sulfonimide (TFSI)), which uniformly enhances the photoluminescence and minority carrier lifetime of monolayer $MoS_2$ by more than two orders of magnitude.[18] Currently, lithographically fabricated MoS$_2$ transistors suffer from two imperfections detrimental to low power CMOS applications: 1.) the large turn-on gate voltages ($V_T$) needed to switch the transistor state and 2.) severely degraded subthreshold swing. The origin of both these imperfections can be traced back to the high density of interface trap charges and substrate interactions. In particular, the large turn-on voltages in MoS$_2$-based FETs arises due to Sulphur vacancies, which are known to inherent *n*-dope the MoS$_2$ channel, moving it away from charge neutrality.

In the work presented here, we show that controlled exposure to a downstream (remotely generated) oxygen plasma dramatically improves in the measured transport (*I-V*) characteristics. The turn-on voltage ($V_T$) moves closer to zero gate voltage due to screening of charged impurities, and a reduction of the sub-threshold voltage swing is observed. Typically, devices fabricated in this study exhibit carrier mobilities (~1-10 cm$^2$/V·s) near typically reported values in literature. On occasion however, a fabricated device may yield rather low mobility. In such rare cases, where the device mobility is remarkably low (~0.01 cm$^2$/V·s), likely limited by impurity scattering, we see a dramatic (up to 190-fold) improvement in carrier mobilities.

We mechanically exfoliate MoS$_2$ onto a pre-cleaned Si/SiO$_2$ wafer. Few-layer (3-15 layers) flakes are located using optical microscopy, and Ti (5nm)/Au (50nm) source and drain electrodes are patterned using electron beam lithography, as shown in Figure 1. The underlying silicon is used to provide a back gate for the sample, as illustrated in the schematic diagram in Figure 1a. Following device fabrication, transport characteristics (i.e., current-gate voltage curves) are measured using a probe station under ambient conditions. The device is then exposed to a remotely generated oxygen plasma as described below, and the transport measurements are repeated on the same device. The plasma used for the cleaning process is remotely generated using ambient air with 20W of RF power, and the sample is placed in a chamber (XEI Scientific

Evactron) at 200mTorr, about 10cm upstream from the plasma source, for 2 minutes. The downstream (or remote) generation of plasma is crucial to minimizing direct bombardment of the MoS$_2$ by ionic species accelerated in the electric field. In this way, one ensures that the removal of impurities and surface contaminants is driven by the chemical reactivity of the oxygen radicals, rather than an unselective physical bombardment process. The air generated plasma, as used in this study, typically comprises of electrons, nitrogen and oxygen ions, as well as neutral oxygen radicals.

In Figure 1c, the source-drain current ($I_D$) is plotted on a log scale as a function of the applied gate voltage $V_G$. The device shows *n*-type conductance at positive applied gate voltages, and is turned off at negative gate voltages. The devices also exhibit a linear dependence of current on the applied source-drain bias voltages (as shown in Figure S1 of the Supplemental Document), which indicates that Ohmic contacts are made with the metal contacts. This particular device shows a shift of the turn-on voltage from -19.4V to -0.9V after exposure to the remote oxygen plasma. For the 8 FET devices measured in this study, we observe a shift in the turn-on voltage from an average of -18.7V to -0.9V, as summarized in Table 1. As mentioned before, the large magnitude of the turn-on voltage arises due to the abundance of sulphur vacancies in mechanically exfoliated MoS$_2$, and this has been reported previously by various groups. Table S1 provides a list of reported values of $V_T$ collected from different experimental studies. The plasma is comprised of charged ions and neutral O radicals that bind readily to S-vacancies and form covalent bonds with the Mo atoms, since O has the same valence number as S, thus mitigating the doping effects associated with the vacancy. In fact, O has a slightly higher electronegativity than S and will bind more strongly than a corresponding S atom.

It should be noted that the results obtained here using a remotely generated oxygen plasma stand in contrast to those obtained utilizing a conventional local O$_2$ plasma. Islam et al. reported the destructive effects of oxygen plasma exposure to the surface of MoS$_2$, showing a severe degradation of the FET mobility after just a few seconds of exposure to the locally generated oxygen plasma[19]. In contrast, using the gentler, remotely generated O$_2$ plasma, we typically observe only a slight degradation of carrier mobility, despite using considerably longer exposure times (approximately 3 minutes).

Our initial intent was to use this oxygen plasma to selectively remove surface residue created during the lithographic process. The AFM images in Figure 2 show an exfoliated MoS$_2$ flake before and after oxygen plasma treatment. Before plasma treatment, this flake exhibits substantial surface residues left over from the lithographic processes (i.e., ethyl lactate-6, PMMA 950 C2, acetone, isoproanol). It is also possible that the tape residue from mechanical exfoliation process also contributes to this unwanted surface contamination on the MoS$_2$ flake. After plasma cleaning, a majority of this residue is removed, without damaging the MoS$_2$ flake itself.

Also evident from the *I-V* characteristics of the Figure 1c, is the reduction of the sub-threshold gate-voltage swing (*S*) after treatment with oxygen plasma. The sub-threshold swing is defined as the change in gate voltage required to reduce the current by one decade, i.e., $S = \log(10) \cdot \frac{dV_G}{d(\log(I_D))}$, where the derivative $dV_G/d(\log(I_D))$ is taken at the onset of current. In conventional FETs, sub-threshold transport is dominated by carrier diffusion, and the sub-threshold swing is largely determined by the MOS capacitance, given by the expression $S = \log(10) \cdot \frac{dV_G}{d(\log(I_D))} \approx \frac{kT}{q}\log(10)\left(1 + \frac{C+C_D+C_{it}}{C}\right)$. Here, $C$ is the oxide (back-gate) capacitance[20], $C_D$ is the depletion layer capacitance, and $C_{it}$ is the capacitance associated with the interface trap charges. From the data in Figure 1, we obtain values of 5.26 V/dec and 3.05 V/dec for the sub-threshold swing before and after oxygen plasma treatment, respectively. This reduction in sub-threshold swing too may be understood as a consequence of the reduction in interfacial charge density, arising primarily due to Sulfur vacancies. The two-fold improvement in sub-threshold voltage swing is important for efficient device operation, as it enables switching the FET state (on/off) at low voltages, and hence minimizes power dissipation.

The field effect mobility of our devices is obtained from the transconductance data shown in Figure 3 using the expression $\mu = \left[\frac{dI_D}{dV_G}\right] \cdot \frac{L}{WCV_{DS}}$, where $W$ and $L$ are the channel width and length, respectively, $C$ is the back gate capacitance, $V_G$ is the gate voltage, $V_{DS}$ is the bias (source-drain) voltage, and $I_D$ is the current flowing through the channel. Here, the derivative $dI_D/dV_G$ is taken at the steepest point on the *I-V* curve from Figure 3. Although this observed device mobility can depend on various factors, such as contact resistance, and surface interactions, looking at Table 1, we observe two distinct trends in the change in mobility due to plasma treatment. For devices showing moderate carrier mobilities (~10 to 100cm$^2$/V·s), the mobility is typically slightly reduced (by approximately a factor of 2) due to the plasma treatment, indicating that exposure to oxygen plasma does introduce some disorder in the MoS$_2$ lattice. However, devices with relatively poor carrier mobilities (~0.001cm$^2$/V·s) show significantly enhanced carrier mobilities after plasma treatment. The low motilities in these devices indicate they suffer from strong extrinsic scattering effects, such as presence of tape residue on the MoS$_2$ surface, rendering them useless. On average, we observe a 28-fold improvement in device mobility for such "damaged" devices, as summarized in Table 1. We attribute this dichotomy in trends to the competing effects of oxygen plasma exposure on the MoS$_2$ surface. While removal of surface impurities from the surface of "damaged" flakes reduces impurity scattering and improves mobility, the direct bombardment of the MoS$_2$ flake by plasma creates disorder, and degrades the carrier mobility. Further, the increase in the "off-state" current, as seen in Figure 1c, is also likely a consequence of an increased density of electronic states within the MoS$_2$ band gap. Here, the use of a remotely generated plasma minimizes the latter effect, allowing the removal of impurities, without significantly increasing the defect density. However, the use of longer etching times (over 4 minutes), or larger RF powers, typically does damage to the MoS$_2$ flake itself, with thinner flakes being more susceptible to degradation.

While the same processing steps were used to fabricate all of the devices in this study, we found that the lowest mobility transistors ($\mu < 0.1$cm$^2$/V·s) show signs of organic residue on the surface, as shown in the AFM image in Figure 2. This indicates that the mobility in these cases is not limited by intrinsic processes (such as defect or phonon scattering in the MoS$_2$), but rather scattering from the extrinsic contaminant species. Thus, these devices show enhanced carrier mobilities due to the removal of surface residue through plasma treatment. It is known that MoS$_2$ transistors are also afflicted by surface contaminants and charged impurities, which dope the material *n*-type[21] and move the turn-on voltage away from zero applied gate voltage. The oxygen plasma consists of negatively charged ions of oxygen and nitrogen, as well as charge neutral free radicals. We believe that these ions and radicals from the plasma bind preferentially to the S-vacancy sites in MoS$_2$, mitigating the doping effects associated with the vacancy. This reduces the net charge on the defect sites, thereby reducing their effect on the turn-on voltage. The optimal etch time of 2-3 minutes corresponds to the amount of time needed to "saturate" all the charged impurity sites. Any further increase in exposure time does not lower the turn-on voltage. In contrast, previous studies report a substantial (100X) reduction in mobilities of MoS$_2$ transistors after even just a few seconds of exposure to locally generated oxygen plasmas[13]. Hence, this comparison demonstrates the relatively gentle nature of the remotely generated plasma as opposed to conventional locally generated plasmas. As listed in Table 1, 3 out of 8 measured devices, show a remarkable increase in carrier mobility by over an order of magnitude. For the particular device shown in Figure 3a, $W=L=4$μm and $C=11.5$nF/cm$^2$ (for a 300nm SiO$_2$ back gate). The estimated mobility of the as-fabricated FET device is 0.017 cm$^2$/V·s, which is considerably lower than typically reported values in literature. However, after treatment with the remote oxygen plasma, the carrier mobility is found to increase to 3.21 cm$^2$/V·s, which is comparable to typical values reported in literature[22]. While most reports in the literature focus on clean FET devices showing higher mobility, our work demonstrates a method to improve even the very worst fabricated devices. Figure 3b shows the $I$-$V_G$ characteristics of another MoS$_2$ device. Here, the carrier mobility drops by 16% from 19.8 cm$^2$/V·s to 16.6 cm$^2$/V·s after the 3 minute oxygen plasma exposure. Interestingly, we observe that exposure to remote oxygen plasma is also accompanied by an increase in the hysteresis of the $I$-$V_G$ characteristics, as shown in Figure S3 of the Supplemental Document. Previous studies on organic transistors[23], inorganic semiconductors[24], as well as carbon nanotubes[25] have shown that hysteresis is a consequence of a dynamic gating of the channel due to mobile surface charges, which may move upon the application of an external field, thereby changing the charge neutrality point. The increased hystereis is similarly attributed to the dynamic gating caused by adsorption of oxygen plasma species onto the MoS$_2$ surface. As reported in our previous work, this oxygen plasma treatment improved the luminescence efficiency of few-layer MoS$_2$ by up to 20-fold due to an indirect-to-direct bandgap transition.[17] In additional, the photoluminescence linewidth becomes substantially narrower after plama treatment, as shown in Figure 4. The photoluminescence spectra allow one to determine the relative lifetime of the excitons, whereas transport measurements give a relative measure of the free carrier lifetimes. In transition metal dichalcogenides, there is a dramatic change in the band structure of the materials, as layer thickness is increased above one monolayer. While the most pronounced aspect of this change is the transition to an indirect gap semiconductor in multilayer TMDCs, this change also has a bearing on the selection rules for allowed intervalley scattering processes in monolayer MoS$_2$.

Hence, the PL spectral linewidth is found to be narrower in monolayer $MoS_2$ than in multilayer $MoS_2$. Since the exposure to oxygen plasma is also shown to decouple individual layers, leading to a monolayer-like band-structure, a similar effect is seen in our spectra This direct gap transition is highly desirable for applications in optoelectronic devices and, coupled with the improved FET device performance, could pave the way for next generation TMDC based devices. It should be noted that the chemical reaction induced by the O-plasma treatment is permanent and stable, and these devices do not revert to their original *I-V* characteristics over time. The plasma treatment, however, does result in an interlayer decoupling, and some delamination of the $MoS_2$ is observed when stored under ambient conditions over the span of a few weeks. Therefore, some strategy for hermetic sealing will have to implemented in order to overcome this instability in practical device applications.

In summary, we demonstrate a reliable and scalable method using an oxygen plasma treatment to dramatically reduce the turn-on voltage required for switching operation in $MoS_2$ transistors and simultaneously improve the subthreshold swing of these devices; both are key parameters for enabling low-power electronic devices and sensing applications. The key novelty in this work lies in bringing $MoS_2$ FETs close to charge neutrality, presumably through the passivation of charged Sulfur vacancies, which typically *n*-dope exfoliated $MoS_2$. Lithographically fabricated FETs also suffer from non-uniformities and huge variability in device-to-device performance, which is greatly reduced upon exposure to the remote oxygen plasma. Our results also shed light on the role of defect and impurity scattering mechanisms limiting device mobilities in $MoS_2$ transistors. While this result is not meant to compete with other methodologies used to make high mobility FET devices, such as using suspended $MoS_2$ or $MoS_2$ sandwiched between two flakes of boron nitride, it provides a scalable route for creating moderate-mobility $MoS_2$ FETs with lower subthreshold swing and turn-on voltages, while significantly reducing the device-to-device variability.

**Acknowledgements:** This research was supported by Department of Energy (DOE) Award No. DE-FG02–07ER46376 (Z. L.) and NSF Award No. 1402906. (R.D).

| Device | $V_T$ (Pre) (V) | $V_T$ (Post) (V) | $\Delta V_T$ (V) | $\mu_{pre}$ (cm$^2$/V·s) | $\mu_{post}$ (cm$^2$/V·s) | $\mu_{post}/\mu_{pre}$ |
|---|---|---|---|---|---|---|
| 1 | -20 | -4 | 16 | 0.017 | 3.21 | 192 |
| 2 | -19 | 4 | 23 | 0.001 | 0.02 | 15.0 |
| 3 | -18 | 1.5 | 19.5 | 0.002 | 0.03 | 18.4 |
| 4 | -18 | 2 | 20 | 19.8 | 16.64 | 0.8 |
| 5 | -18 | -2 | 16 | 25.4 | 21.00 | 0.8 |
| 6 | -22 | -6 | 16 | 26.6 | 12.50 | 0.5 |
| 7 | -20 | -3 | 17 | 25.5 | 15.65 | 0.6 |
| 8 | -15 | 0 | 15 | 18.0 | 14.31 | 0.8 |
| Avg | -18.7 | -0.9 | **17.8** | 14.4 | 10.3 | **28.6** |

Table 1: The FET carrier mobilities (μ) and turn-on voltages ($V_T$) for 8 MoS$_2$ FET devices before and after remote oxygen plasma treatment.

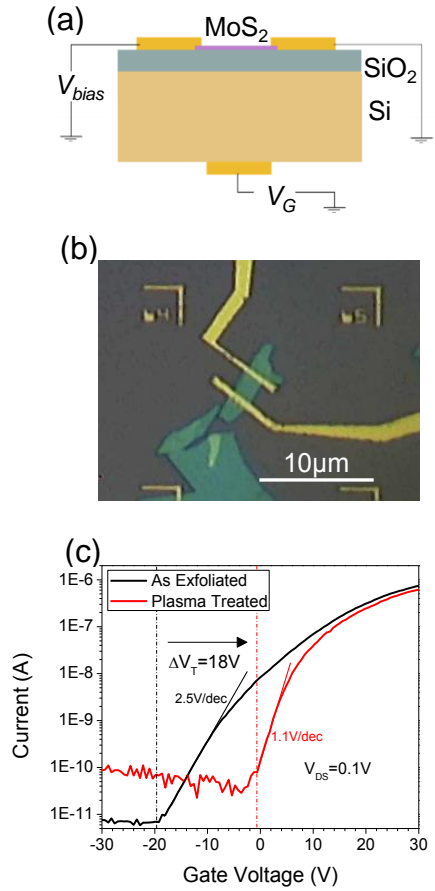

FIG. 1: (a) Schematic diagram and (b) optical microscope image of a back-gated MoS$_2$ FET device. (c) A log-linear plot of the source-drain current as a function of applied gate voltage.

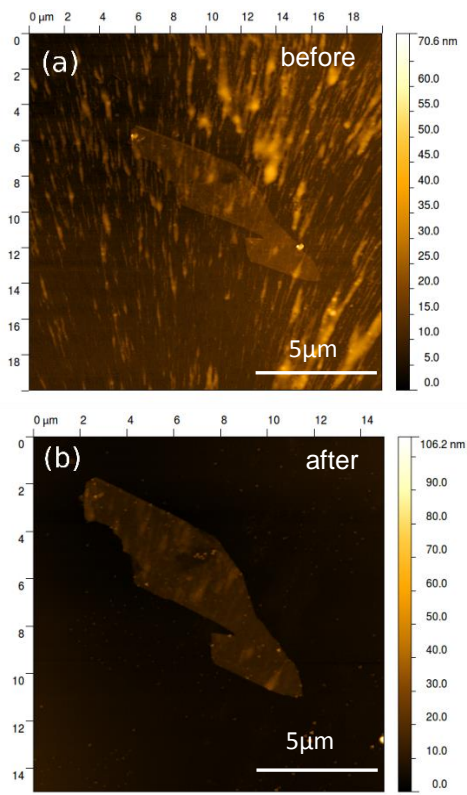

FIG. 2: AFM images of a typical $MoS_2$ flake (a) before and (b) after exposure to remotely generated oxygen plasma.

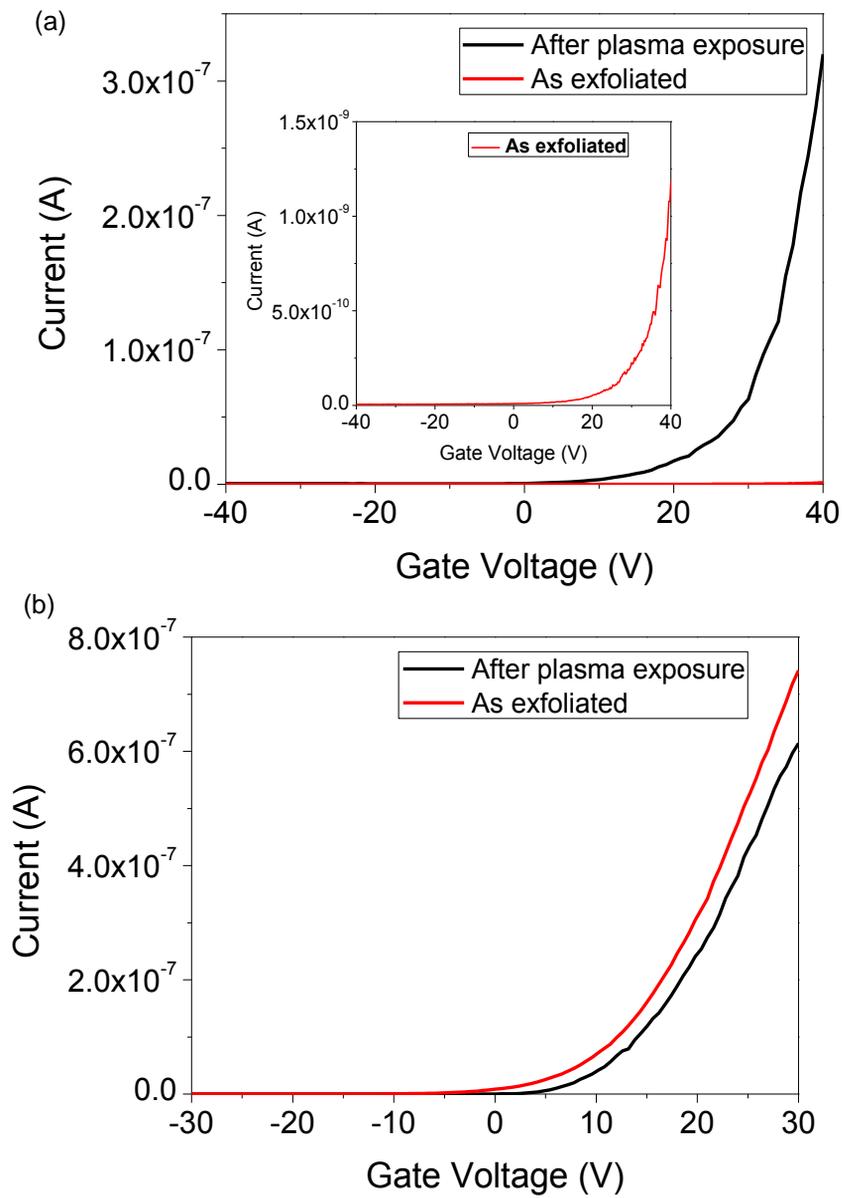

FIG. 3: (a) Current–gate voltage characteristics of a device showing a significant (200X) increase in the field effect carrier mobility after a 3 minute exposure to $O_2$ plasma. The inset shows the *I-V* curve before plasma exposure. (b) *I-V* curve of another device showing a 15% drop in carrier mobility after a three minute exposure to $O_2$ plasma.

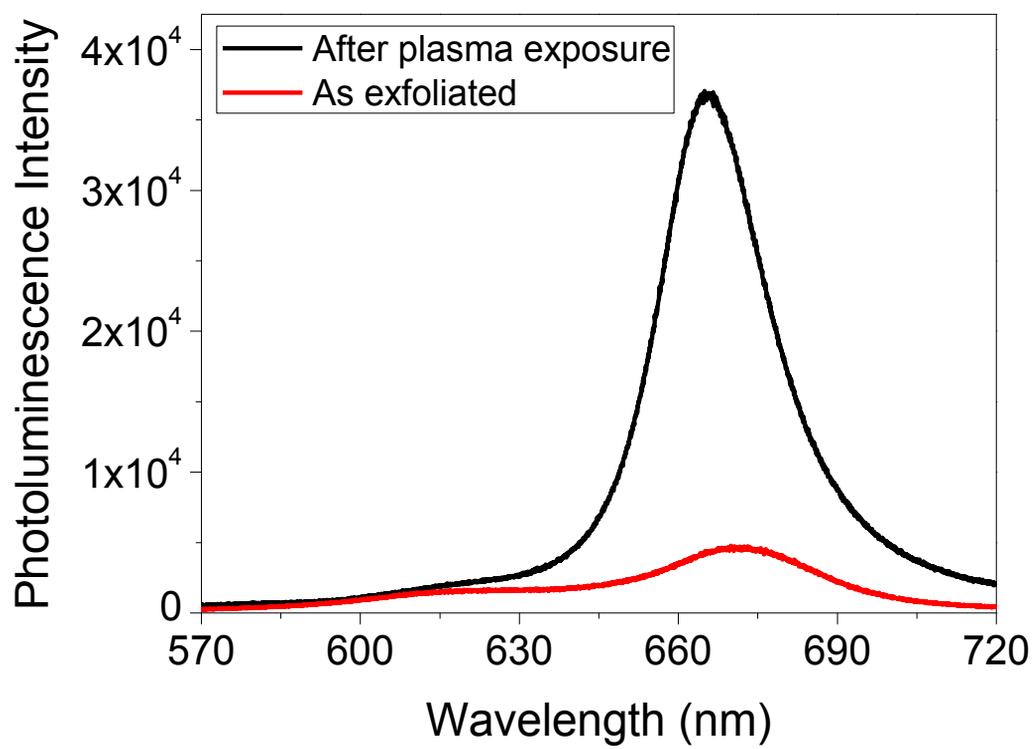

FIG 4: Enhanced photoluminescence emission from few layer $MoS_2$ after exposure to remotely generated oxygen plasma.

# Supplemental Information for
# Charge Neutral MoS$_2$ Field Effect Transistors Through Oxygen Plasma Treatment

Rohan Dhall[1], Zhen Li[1], Ewa Kosmowska[2], Stephen B. Cronin[1]

[1]Ming Hsieh Department of Electrical Engineering, University of Southern California,

Los Angeles, 90089

[2]XEI Scientific, Redwood City, CA 94063

**Ohmic Contacts:**

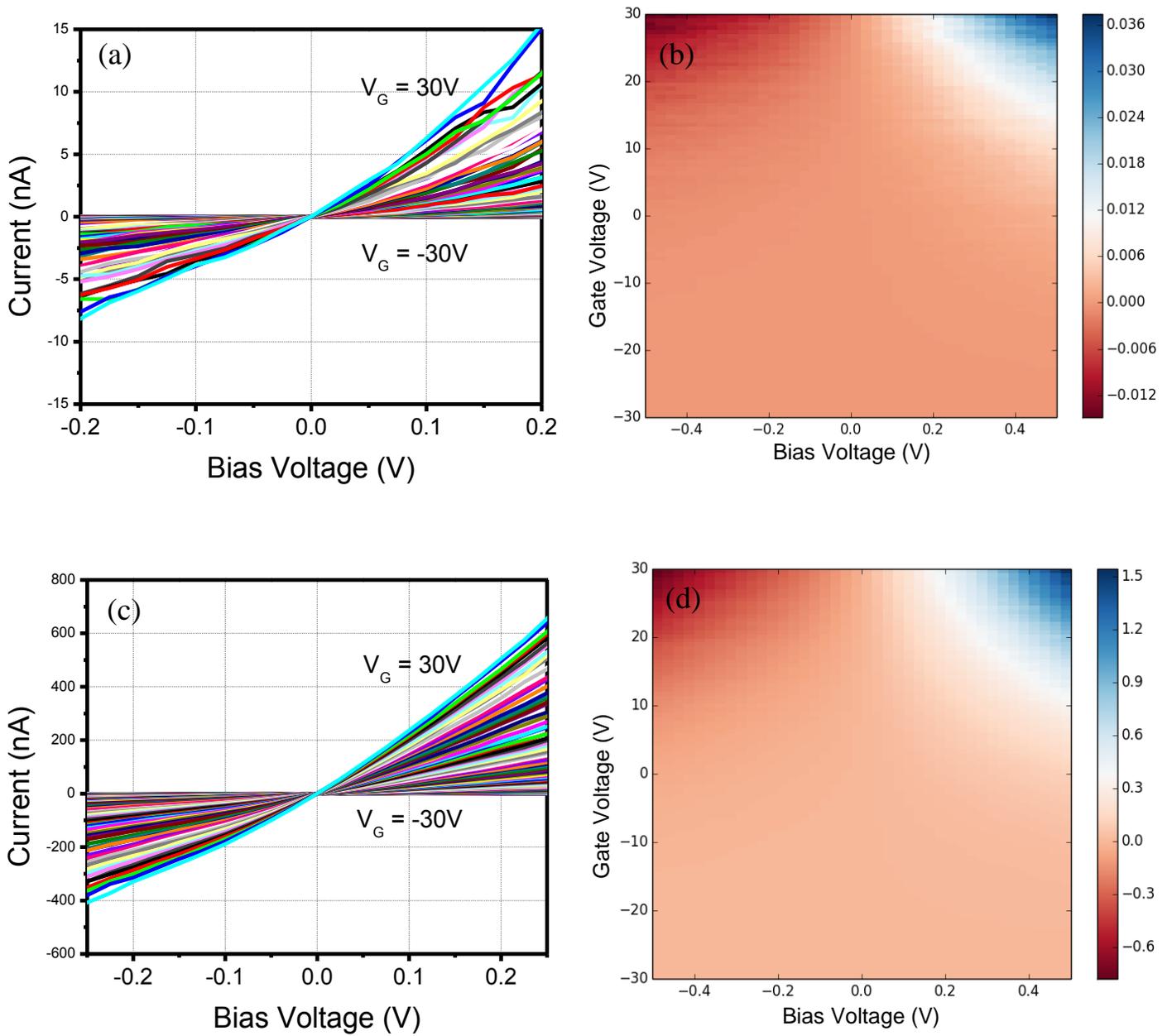

Figure S1: (a) and (c) show the *I-V* characteristics of typical MoS$_2$ FETs showing the formation of a reasonable Ohmic contact to MoS$_2$ with the Ti/Au contacts evaporated for the source and drain electrodes. The gate dependence of the *I-V* curves are illustrated in the color plots in (b) and (d).

Since current through TMDC FETs is often dominated by transmission of carriers through the metal/semiconductor junction, transport characteristics are very sensitive to any change in the band alignments of the metal contacts and semiconductor ($MoS_2$). The underlying physics governing device behavior has been well explained in the context of carbon nanotube Schottky barrier FETs. In order to verify that the observed shift in threshold voltage and reduction of sub-threshold swing are not artifacts due to changes in Schottky barrier heights, we performed scanned photocurrent microscopy using a 532 nm excitation laser, as shown in Figure S2. The outline of the flake is shown by the yellow dashed lines, and the source and drain electrodes are indicated with red dashed lines. In the as-fabricated FET devices, the photocurrent is generated at both metal contacts, which form Schottky junctions[1, 2]. Applying a bias voltage, $V_{DS}$, between the source and drain electrodes, we can modulate the barrier heights and selectively enhance photocurrent generation at either the source or drain by effectively creating a flat band condition at the other electrode. We found that both before and after exposure to oxygen plasma, this condition was achieved by applying a $V_{DS}=\pm200$mV. This experiment leads us to believe that a change in the barrier transmission at the electrodes is not the underlying mechanism for the observed enhanced mobilities and shift in turn-on voltage.

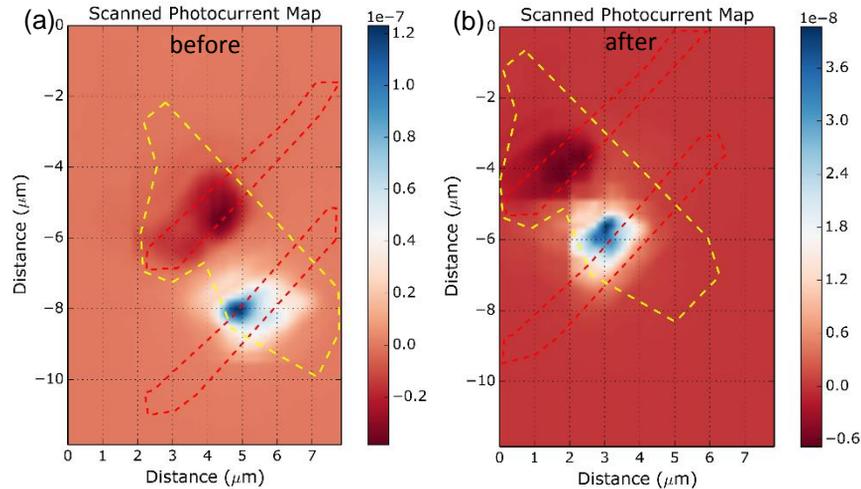

FIG. S2: Scanned photocurrent maps for two-terminal $MoS_2$ devices (a) before and (b) after exposure to oxygen plasma, which shows that the generated photocurrent is dominated by the Schottky barrier at the metal-semiconductor junction and remains relatively unchanged after oxygen plasma treatment.

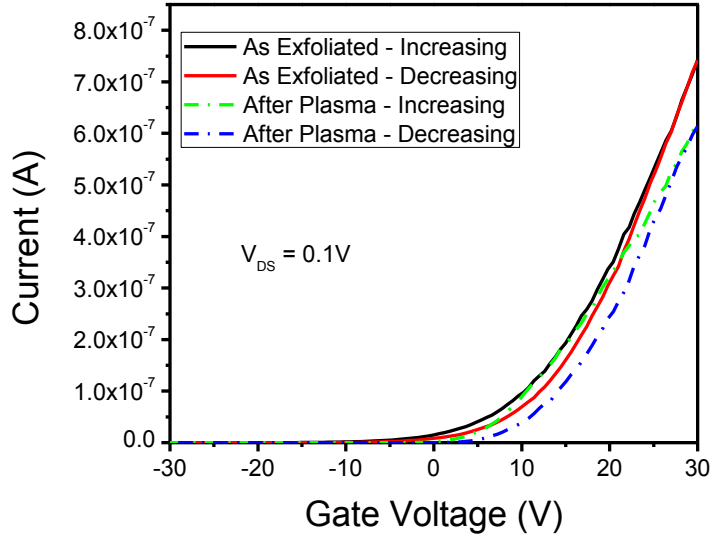

FIG. S3: Current-gate voltage curves for a typical $MoS_2$ FET device, recorded in the forward and reverse sweep directions. After plasma treatment, devices tend to show larger hysteresis, as evident from these plots.

| Gate Oxide (thickness in nm) | Threshold Voltage ($V_T$) | Reference |
|---|---|---|
| $SiO_2$ (100nm) | -30V | [3] |
| $SiO_2$ (270nm) | -4V | [4] |
| $SiO_2$ (300nm) | -60 to -40V | [5] |
| $SiO_2$ (90nm) | -4.5V | [6] |
| $SiO_2$ (300nm) | -7.5V | [7] |
| $SiO_2$ (270nm) | -3 to 4 V | [8] |
| $Y_2O_3/HfO_2$ stack | -3 to 4 V | [9] |

TAB. S1: Turn-on voltages for molybdenum disulfide field effect transistors reported by various different studies.